# Single Whispering-Gallery-Mode Resonator With Microfluidic Chip as a Basis for Multifrequency Microwave Permittivity Measurement of Liquids

Alexey I. Gubin, Irina A. Protsenko, Alexander A. Barannik, Nickolay T. Cherpak, *IEEE Senior Member,* Svetlana A. Vitusevich

***Abstract* — The accurate measurement of complex liquid permittivity in a frequency range provides important information on liquid properties in comparison to single frequency permittivity investigations. The multifrequency microwave characterization technique based on a single quartz whispering-gallery-mode (WGM) resonator with a microfluidic chip are proposed, developed, and demonstrated. This technique allows the complex permittivity of small volumes of liquid to be measured at six resonant frequencies in the 30–40 GHz frequency range. The calibration is performed by simulating the measurement cell and by plotting the calibration nomogram charts for all six investigated frequencies. The novel approach is applied in studies of the complex permittivity of the L-lysine in water solutions. The results open up possibilities to investigate the complex permittivity of biological liquids at several frequencies and to further develop the microwave dielectrometry of small liquid volumes in a certain frequency range using the quasi-optical nature of a single high-Q WGM resonator.**

## I. Introduction

THE investigation of liquids is important in many areas of science and technology. One of the main characteristics of liquids is complex permittivity, which allows the liquid to be identified, the concentration of the substance in a water solution to be determined, and different processes to be investigated, for instance. Liquid permittivity measurement techniques can be divided into two subgroups: resonance and non-resonance techniques. Resonance techniques are developed on the basis of utilizing different resonant structures with a small amount of a liquid, which results in a change of the resonant characteristics of the whole structure [1],[2]. Non-resonance techniques use the parts of different transmission lines and also utilize an amount of liquid [3]-[5]. The liquid causes a change in the transmission/reflection characteristics of the structure. Resonance techniques are more sensitive than non-resonance ones. However, as a rule, they are applied to measure liquids at a single frequency, while non-resonance techniques allow measurements in a certain, sometimes wide, frequency range [6]. In general, high sensitivity is more important for investigations of different processes in liquids, while measurements in a wide frequency range are more important for liquid identification. However, measurements can also take place in the opposite situation, when a resonant technique is utilized to identify liquids using two significantly separated frequencies excited in two different resonators [7] and a broadband technique is applied to study physical or biophysical processes [8].

The dielectric whispering-gallery-mode (WGM) resonator-based microwave technique demonstrates high sensitivity and accuracy for the characterization of a liquid being tested [2]. This is determined by the high quality factor (Q factor) of such resonators. A high sensitivity and accuracy of WGM resonator-based methods has been achieved in the optical range [9],[10] as well as in the microwave region [2],[11]-[15]. The investigated object placed in the resonant structure changes its resonant frequency and Q factor. Typically, the object position is determined in relation to the electromagnetic field pattern to maximize the sensitivity of the resonance parameters to the permittivity of the object under investigation. There are a number of ways to introduce the liquid being tested to the microwave resonant WGM structure: using a two-layer resonator [11], a droplet on the resonator surface [12], a capillary placed in the hole of the resonator [13],[14], or a microfluidic channel placed on the resonator [2]. The WGM resonator with microfluidic chip technique has several advantages, such as the closed microfluidic system with constant geometrical dimensions (i.e. no evaporation, as in the case of the droplet on the resonator surface) and the use of the

A. I. Gubin, I. A Protsenko, A. A. Barannik, and N. T. Cherpak are with the Department of Solid-State Radiophysics, O. Ya. Usikov Institute for Radiophysics and Electronics, NAS of Ukraine, 61085 Kharkiv, Ukraine (e-mail: gubin@ire.kharkov.ua).

S. Vitusevich is with the Institute of Biological Information Processing - Bioelectronics (IBI-3), Forschungszentrum Jülich, 52425 Jülich, Germany (corresponding author; e-mail: s.vitusevich@fz-juelich.de).

The authors would like to thank the German Research Foundation (DFG), project VI 456/4, for their financial support.

dielectric resonator without its mechanical modification (i.e. drilling a hole for the capillary). The relative accuracy of the permittivity measurement of the glucose aqueous solutions obtained using the technique based on a sapphire WGM resonator with a microfluidic chip is about 1.4% for the real part and 0.7% for the imaginary part [2]. When using a quartz WGM resonator, the accuracy is even better than in the case of sapphire. It is about 0.7% for the real part and 0.4% for the imaginary part [15]. Liquid investigations were performed exclusively at a single frequency of about 35 GHz.

However, the frequency dependence of the permittivity liquid often provides additional information, such as the identification of the liquid under investigation and an understanding of the relaxation mechanisms. Usually, such measurements are performed using a non-resonant technique [8],[16],[17]. Non-resonance methods that are currently used for measurements at low liquid volumes have proper calibration schemes, but the difference in scattering parameters to be measured is low (less than 0.01 dB in [16]). The investigation of small liquid volumes by resonance methods does not require highly precise scattering parameter measurements to achieve a high determination accuracy for resonant parameters. The measurements on dual resonances were performed in the optical range [18]. In principle, it is possible to perform a similar investigation in the microwave region. The main disadvantage of such measurements is the spacing between the eigenfrequencies of the resonators at the lowest modes. As a rule, these measurements are generally not acceptable for research purposes. When using WGM resonators, it is possible to perform investigations for a number of modes with significantly lower frequency spacing (about 2 GHz in our study). Moreover, the coupling must be tuned to become symmetrical to the resonance line in order to analyze it with a -3 dB level frequency bandwidth or with a Lorentzian function fitting. Changing the coupling during the single object study requires additional time and increases the measurement error. In this respect, the technique has to be developed further to study liquids with a high level of accuracy at several resonant frequencies in a certain frequency range using the same single WGM resonator.

In this paper, we present the results of the development of a single WGM resonator with microfluidic chip for measurement**s** at six resonant frequencies (resonant modes) without need for the coupling to be tuned. We show that in the case of using WGM resonators fabricated from crystal quartz material, it is possible to perform investigations for a number of modes with spacing of about 2 GHz. The resonance structure was modeled using the COMSOL Multiphysics software [19]. The six calibration nomogram charts were plotted for the measurements at each of the modes. The calculation of nomogram charts takes time, but it should be performed only once and prior to the experimental investigation. It does not therefore increase the measurement time. The resonance frequency shift and changes of the inverse Q factor were extracted by fitting the measured frequency responses for different concentrations of lysine in water solutions. The dependencies of the complex permittivity of solutions on lysine concentration were obtained using such a calibration with nomogram charts for all six modes. This corresponds to measurements at frequencies of 30.16 GHz, 31.99 GHz, 33.83 GHz, 35.66 GHz, 37.5 GHz, and 39.34 GHz, and takes into account the frequency dependence of the single crystal quartz permittivity. It should be emphasized that lysine is a basic aliphatic amino acid. It is used in protein synthesis. It plays an essential role for humans and the human body cannot synthesize it. There are a number of papers dedicated to investigating the microwave permittivity of L-lysine in water solutions using non-resonant methods [17],[20],[21].

## II. Measurement technique

The measurement technique was developed based on experience and the results obtained using dielectric WGM resonators with microfluidic chip (MFC) and capillary. It was successfully used for complex liquid permittivity measurements at a single resonant frequency [2],[13]-[15],[22]. The quasi-optical nature of the single WGM resonator allows, in principle, the use of a multi-resonant approach in the technique for the microwave characterization of material media with acceptable spacing between the resonant frequencies. All of the previous measurements were performed exclusively on a definite mode, as the measurement**s** on a number of modes/frequencies require the coupling to be tuned at almost every resonant frequency. This is important for two reasons: i) to obtain a small enough coupling and ii) to obtain a symmetrical resonance line. The latter was commonly accepted for extracting accurate resonator parameters, as the -3 dB extraction method and the method with the Lorentzian fit are only accurate in the case of a symmetrical line. However, we reveal experimentally that it is possible to extract resonator parameters for six multimode measurements using the single resonator with the same coupling. We show that the fitting of experimentally obtained data using the applied model enables resonator parameters to be extracted for six multimode measurements using the single resonator with the same coupling in a relatively wide frequency range (10 GHz), as will be shown below.

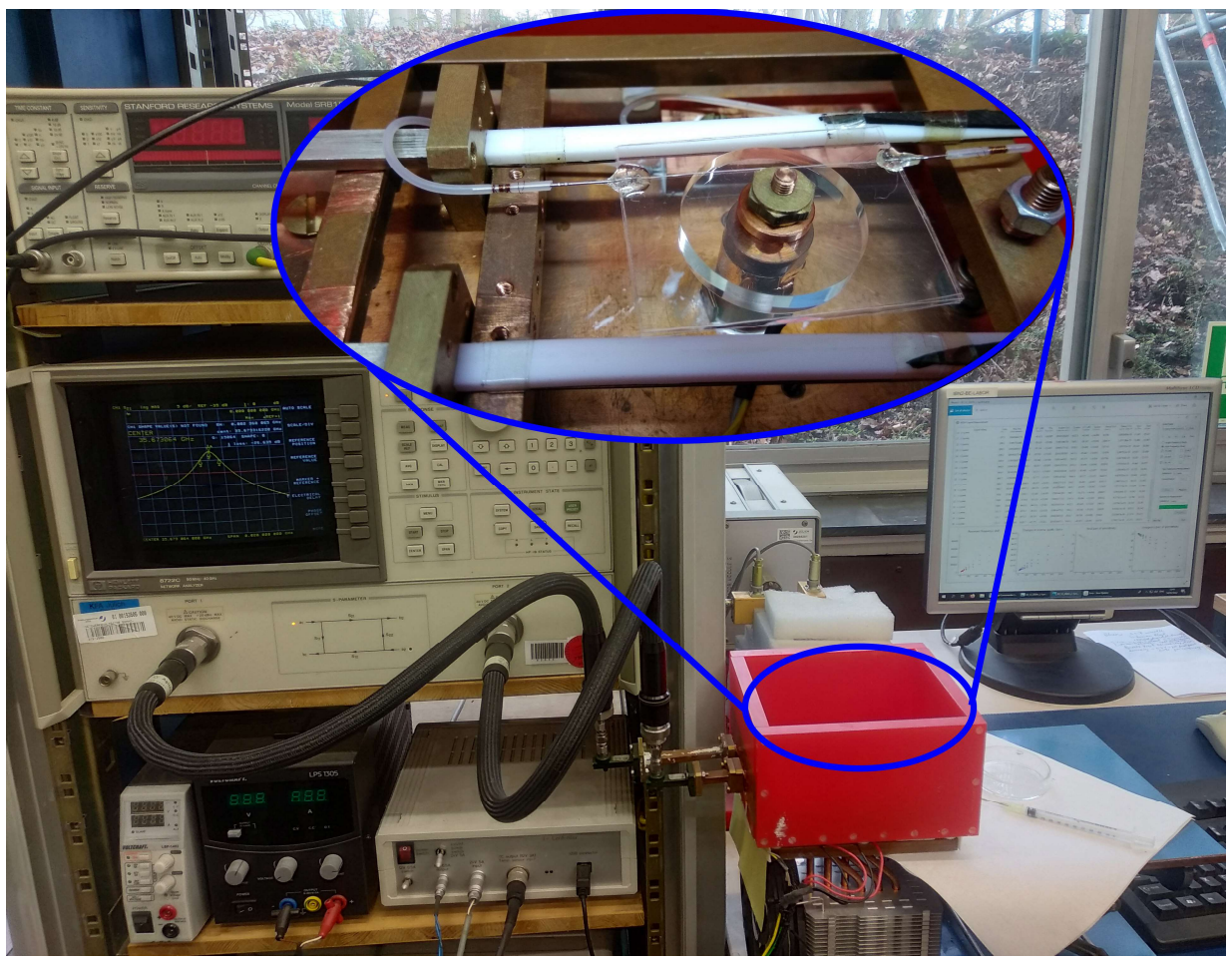

Fig. 1. Experimental setup consisting of single crystal quartz WGM resonator measurement cell in an isolated red box connected to a network analyzer (by coaxial to waveguide transition) and temperature controller.

*A. Experimental setup*

The resonator technique is developed on the basis of the WGM resonator fabricated from single crystal quartz material with MFC (Fig.1). The pair of mirror dielectric waveguides (one input and one output) excited by metal waveguides are connected by coaxial to waveguide transitions with the network analyzer. The material of the dielectric waveguides is Teflon and the metallization is made from copper. The setup is placed in an isolated box with a temperature stabilization of 0.01 °C [15],[22]. The metal with a low thermal conductivity coefficient (nickel silver) and small thickness was used as a metal waveguide material to minimize the heat transfer from the outside space. The MFC dimensions are 30×40×0.73 mm [Fig. 2(a)]. It is fabricated from low-loss ZEONOR 1420 plastic [23]. The microfluidic channel with a diameter of 0.21 mm is placed at the midway point of the plastic height as a result of using the thermocompression method. The volume of the channel to be filled by the liquid under investigation is about 1.2 μl, but its volume interacting with the resonator electromagnetic field is lower (about 0.4 μl). The stainless steel tubes are sealed at the channel edges to fill the liquid under investigation. The channel obtained using this procedure is a smooth cylindrical tube [Fig. 2(b)]. The hole is drilled in the center of the MFC to fix it and place it in the center on top of the quartz resonator.

The $S_{21}$ response was recorded by the 8722C network analyzer for a number of resonance modes. The measurements were performed using six modes $HE_{m\,1\,1}$, where m=12, 13, 14, 15, 16, and 17**,** which correspond to frequencies of about 30.16 GHz, 31.99 GHz, 33.83 GHz, 35.66 GHz, 37.5 GHz, and 39.34 GHz respectively. The modes were identified by investigating the bare resonator in [15]. The frequency responses were recorded in a narrow band around the six resonant frequencies selected in the frequency range from 30 GHz to 40 GHz. The difference between the adjacent resonant frequencies is mainly determined by the diameter of the disk resonator [24]. By altering the diameter, it is possible to change the values of resonant frequencies, their spacing, and the resulting number of modes of such a multi-resonant approach, which is important for different measurement requirements.

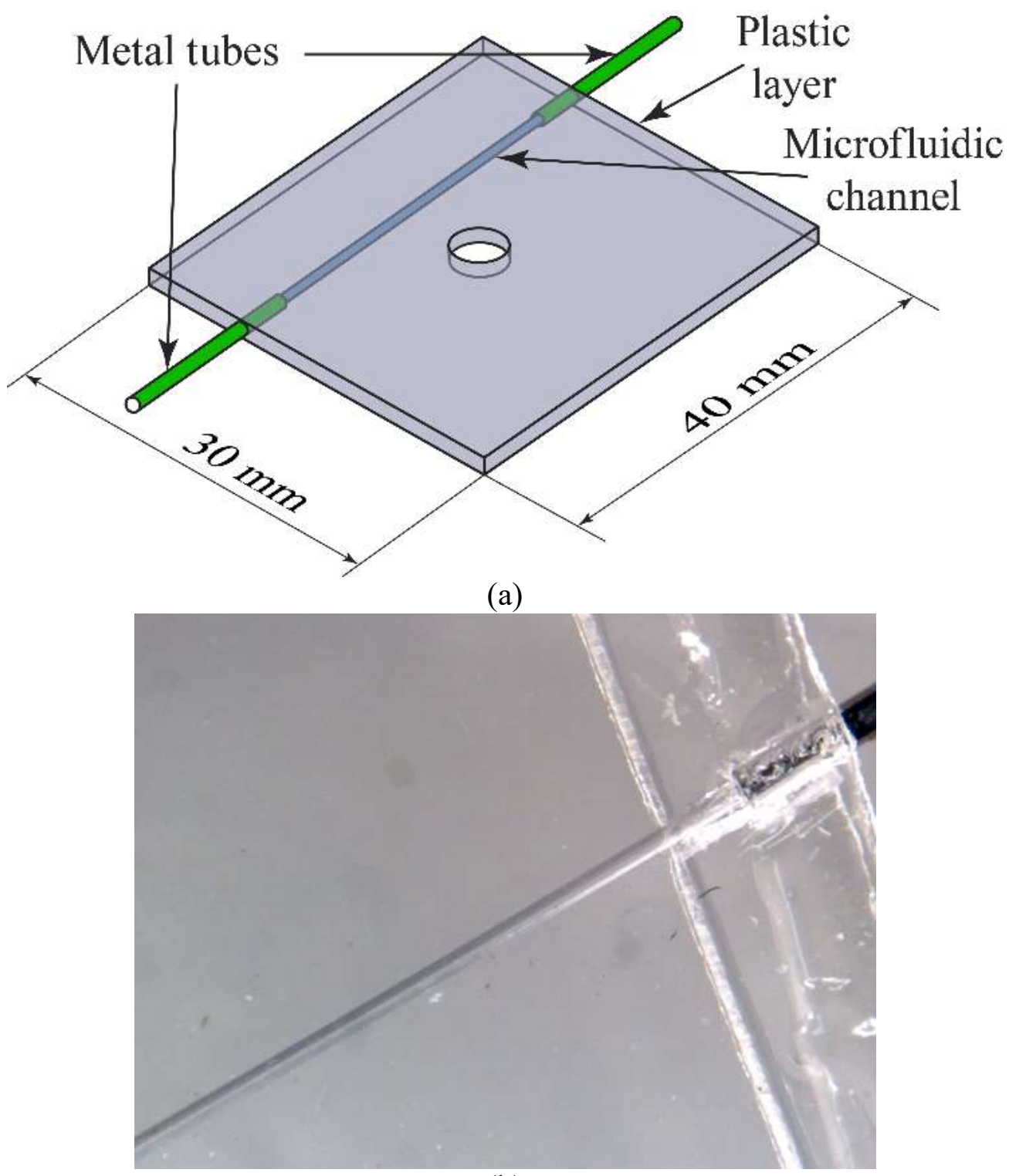

(a)

(b)

Fig. 2. Microfluidic chip (MFC): (a) schematic and (b) microfluidic channel microscope photo.

*B. Extracting resonator parameters*

The investigation of complex liquid permittivity for a number of frequencies/modes without changing the coupling is fast and convenient. Extracting the Q factor and resonance frequency using common methods such as the Lorentzian fit and the -3 dB level method is only accurate for symmetrical resonance frequency responses. Unfortunately, it is not possible to tune the coupling to reach all six resonances so that it is symmetrical. It is difficult to perform even for two resonance modes. Controlling the coupling using some mechanical elements driven outside the isolating box or, in particular, its opening requires too much time (comparable to the measurement time and sometimes even greater). In this respect, instead of fitting using the Lorentzian method, we suggest using the coupled modes parameter extraction approach [25]. It is based on an approximation of the amplitude-frequency response of the resonator by a squared sum of the complex amplitudes of the mode. In our particular case, each single nonsymmetrical resonance can be fitted using the following complex function:

$$G(f) = G_s + \frac{A_1 e^{i\varphi_1}}{1+2iQ_1(f-f_1)/f_1}, \qquad (1)$$

where $G_s$ is the real number, $\varphi_1$ is the phase difference, $A_1$ is the real coefficient, $Q_1$ is the quality factor, $f_1$ is the resonance frequency, and $f$ is the current frequency.

The responses of the measurement cell with the channel filled by 1 mol/l lysine in water solutions are shown in Fig.3.

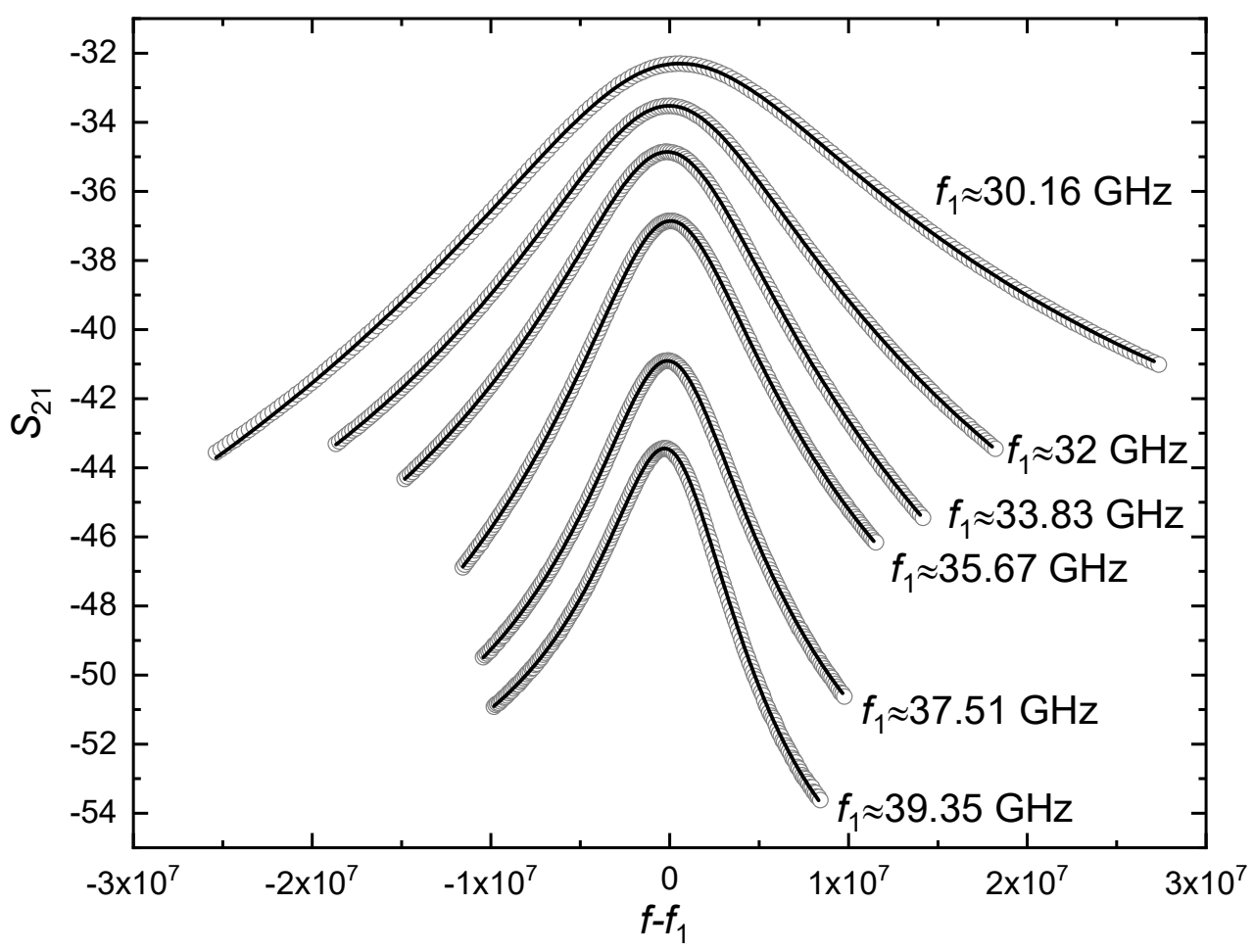

Fig. 3. Responses of the measured (dots) and fitted (lines) by Eq. (1) in the case of microfluidic channel filling by 1 mol/l lysine in water solution. Resonance frequencies $f_1$ are equal to 30160307413 Hz; 31994880988 Hz; 33830950246 Hz; 35668512932 Hz; 37507731954 Hz; and 39348380525 Hz.

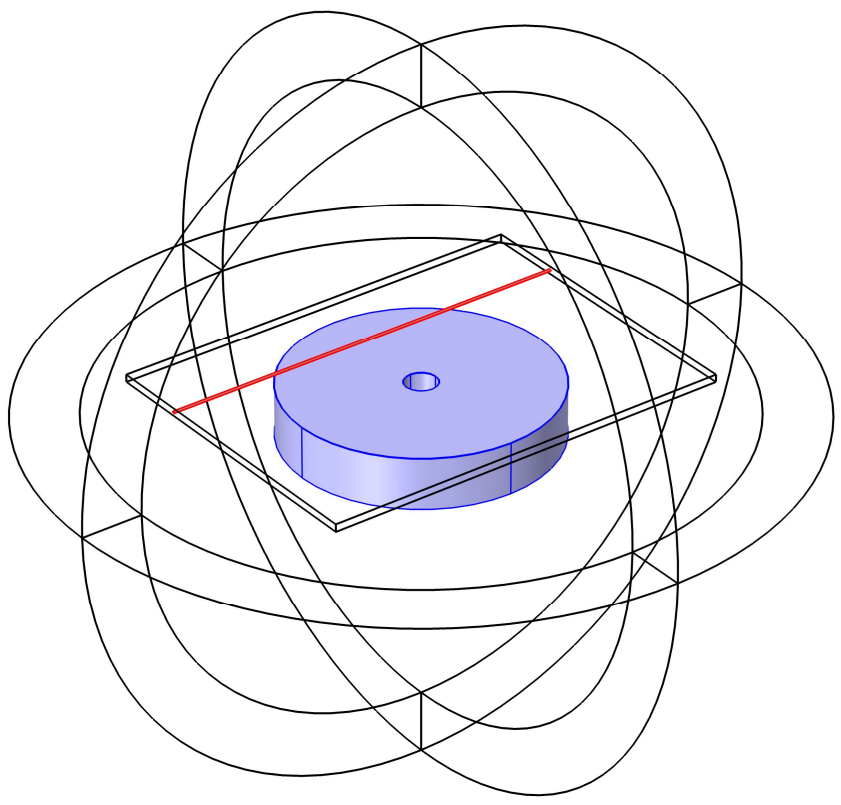

(a)

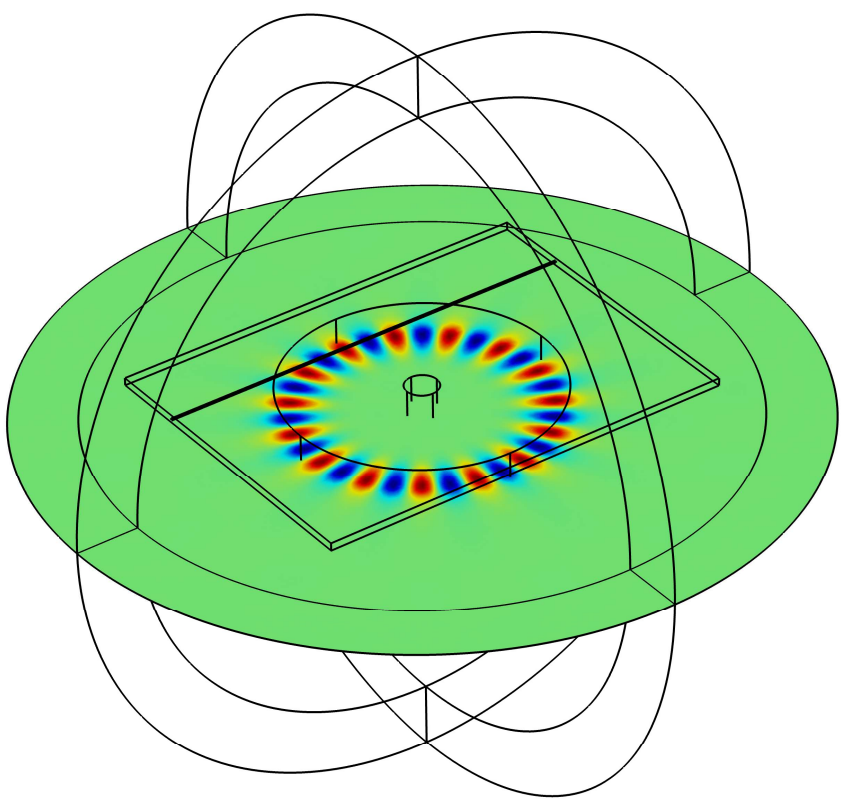

(b)

Fig. 4. Quartz WGM resonator with MFC: (a) the model realized in COMSOL with the quartz disk resonator and (b) its electric field pattern.

The resonance frequencies for six modes are obtained to be equal to 30160307413 Hz; 31994880988 Hz; 33830950246 Hz; 35668512932 Hz; 37507731954 Hz; and 39348380525 Hz. The function according to Eq. (1) perfectly fits the nonsymmetrical resonance responses (solid lines in Fig.3). The standard deviation error of the resonance frequency determination varies from 1 kHz ($3\cdot10^{-6}$ %) to 5.4 kHz ($1.8\cdot10^{-5}$ %) and the Q factor determination varies from 0.6 (0.03%) to 5.1 (0.09%).

Using this procedure allows us to perform measurements of a liquid to be tested on six modes without needing to tune the coupling. We extracted the resonant parameters of six modes in a frequency range of about 10 GHz using the above-described fitting approach. The fitting procedure was realized using Python code.

It should be noted that as a material used to produce the resonator, single crystal quartz not only offers higher measurement accuracy and sensitivity compared to a sapphire resonator, but also a weaker dependence of the coupling coefficient of transmission lines with a resonator due to the noticeably lower value of the real part of the quartz permittivity (~ 4) compared to sapphire (~ 10). The latter contributes to the development and implementation of the multi-resonance measurement technique based on WGM resonators.

## C. Modeling results

The simulations of the measurement cell were performed by COMSOL Multiphysics [19]. To obtain complex resonance frequency values, the eigenvalue solver was used. Such an approach allows the real structure to be described appropriately, as the distance between the resonator surface and waveguides is long enough to minimize the coupling influence on the resonator characteristics. The resonator model includes the rectangular (30×40 mm$^2$) plastic chip with the microfluidic channel [schematically represented in Fig. 2(a) and microscopic image Fig. 2(b)] placed onto the quartz disk resonator [Fig. 4(a)]. The resonator diameter and the height are equal to 24.99 mm and 4.95 mm respectively. The inner hole diameter is equal to 3.05 mm. The channel diameter is difficult to measure exactly but is estimated to be about 0.21 mm. In order to precisely define the values of the channel diameter and its location in the plastic layer, these values are changed in the model within the measurement margins of error and the obtained calculated data are compared with the experimental data in the whole frequency range. The spherical scattering boundary in combination with the perfectly matched layer is used to imitate the open space around the resonator and to exclude the reflection phenomena on the sphere boundary.

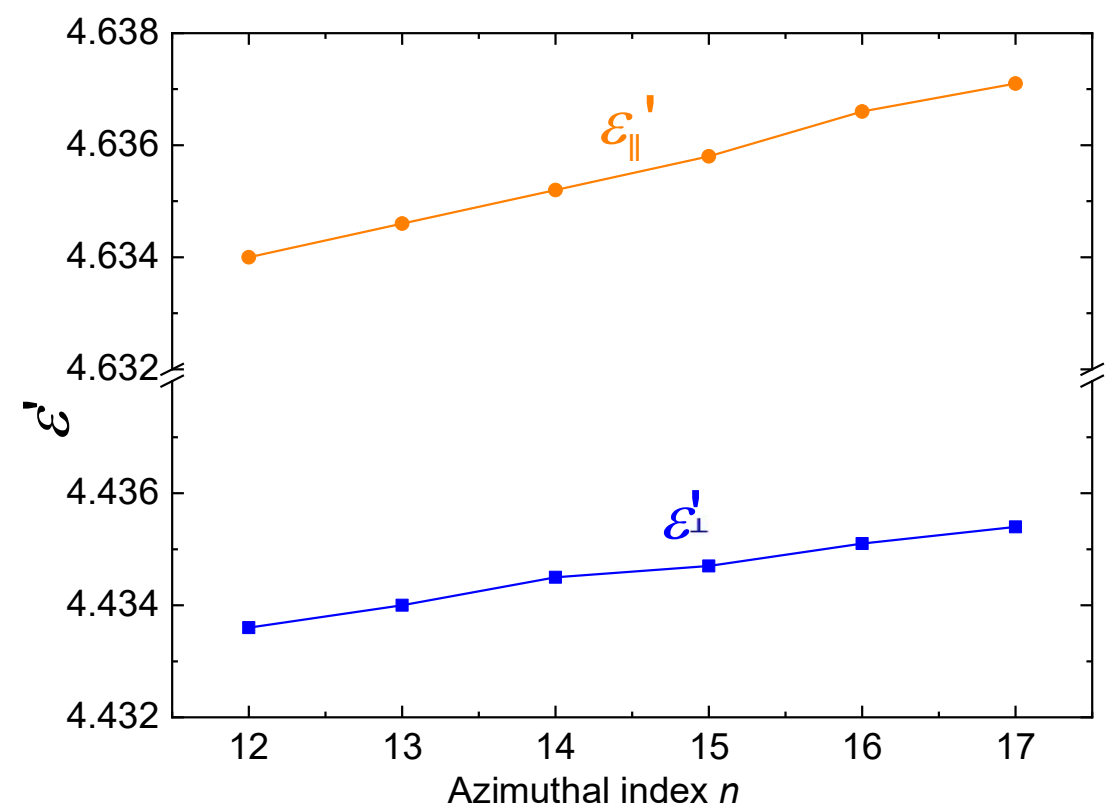

(a)

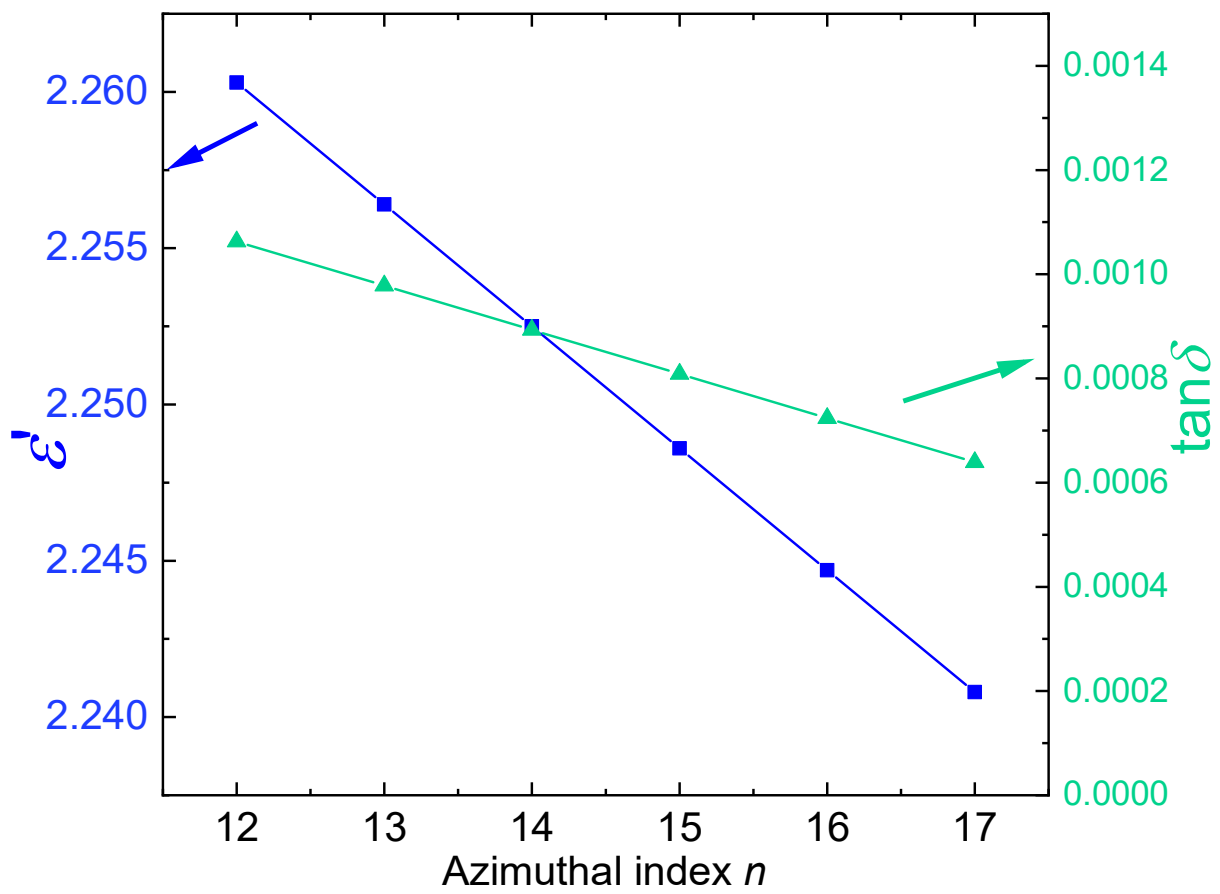

(b)

Fig. 5. Effective values of the electrophysical parameters of (a): quartz and (b): plastic for modes with different azimuthal indices obtained using the procedure described in the text.

Fig.4(b) demonstrates the field pattern of whispering gallery mode $HE_{15\,1\,0}$ in the quartz resonator. This mode is chosen as an example; for other modes, the azimuthal index *n* is different and the higher the *n* value, the closer that the field pattern concentrated to the resonator cylindrical surface is. The microfluidic channel is partially placed in the electromagnetic field so that the liquid parameters have an influence on the resonator characteristics. The model allows an investigation of the mode splitting caused by the non-homogeneous violation of the axial symmetry of the resonator. The frequency difference between the mode peaks is too small (about 20 kHz) to separate these modes in the experiment. Therefore, the averaged values of the frequencies and *Q* factors are presented in the paper to compare the calculated data with the experimental data.

The first step of the simulation was to investigate the bare (without the plastic layer) quartz resonator. The properties of single crystal quartz as an anisotropic material are described by a diagonal tensor as follows:

$$\varepsilon = \begin{bmatrix} \varepsilon'_{\perp}(1 - i\tan\delta_{\perp}) & 0 & 0 \\ 0 & \varepsilon_{\perp}(1 - i\tan\delta_{\perp}) & 0 \\ 0 & 0 & \varepsilon_{\parallel}(1 - i\tan\delta_{\parallel}) \end{bmatrix},$$

where $\varepsilon_{\perp}$, $\varepsilon_{\parallel}$, $\tan\delta_{\perp}$, and $\tan\delta_{\parallel}$ are the tensor components that describe the permittivity and loss tangent of quartz in two directions: perpendicular and parallel to the crystal axis ***c***, respectively.

A preliminary calculation of the bare resonator characteristics was presented in [15]. For the calculation, the crystal quartz permittivity values obtained in [26] were used regardless of their dependencies on frequency within the operating range. Since the measurements in [15] were performed exclusively on a single frequency/mode, the permittivity frequency dependence did not need to be taken into account. With the aim of performing the permittivity measurement for multiple modes of this range and enhancing measurement accuracy, it is important to specify the resonator permittivity values as a function of the frequency.

In order to take crystal quartz permittivity dependence into account and hence to improve the measurement accuracy, the permittivity values obtained in [26] are corrected slightly using the following procedure. The values of the perpendicular and parallel permittivity components are specified using a complex of iterations to superpose the experimental and calculated frequency values for HE and EH modes for the whole operating frequency band [Fig. 5(a)]. The values change from 4.4336 to 4.4354 for perpendicular permittivity components and from 4.634 to 4.637 for parallel permittivity components. Loss tangent values should remain constant in the operating frequency region and have been determined on the basis of the measured Q factor values as $\tan\delta_{\perp}$=4.15×$10^{-5}$ and $\tan\delta_{\parallel}$=1.772×$10^{-5}$. The calculated values of the frequency and Q factors using the permittivity and loss tangent corrected in this way are shown in TABLE I. The results of measurements and preliminary calculations from [15] are also shown. It should be noted that the calculated values obtained by considering the specified permittivities dependent on the frequencies describe the real structure more accurately. The differences between measured and simulated resonance frequencies amount up to 3.3·$10^{-3}$ %, taking into account the aforementioned dependence, and up to 9.3·$10^{-2}$ % for preliminary calculations. The differences in the Q factor do not change significantly.

TABLE I
EXPERIMENTAL SPECTRUM OF THE BARE QUARTZ RESONATOR WITH THE MODES IDENTIFIED BY THE SIMULATIONS

| Mode | Simulation [15] | | Simulation (using corrected permittivity values) | | Experiment [15] | |
|---|---|---|---|---|---|---|
| | *f* (GHz) | *Q* | *f* (GHz) | *Q* | *f* (GHz) | *Q* |
| $HE_{12\,1\,1}$ | 30.475 | 6307 | 30.461 | 6379 | 30.462 | 6491 |
| $EH_{12\,1\,1}$ | 31.427 | 5227 | 31.456 | 5178 | 31.456 | 4151 |
| $HE_{13\,1\,1}$ | 32.288 | 13294 | 32.274 | 13440 | 32.274 | 13354 |
| $EH_{13\,1\,1}$ | 33.385 | 9729 | 33.416 | 9635 | 33.416 | 8202 |
| $HE_{14\,1\,1}$ | 34.105 | 23888 | 34.089 | 24101 | 34.089 | 23810 |
| $EH_{14\,1\,1}$ | 35.335 | 15222 | 35.368 | 15119 | 35.368 | 15086 |
| $HE_{15\,1\,1}$ | 35.926 | 34988 | 35.909 | 35197 | 35.908 | 34987 |
| $EH_{15\,1\,1}$ | 37.277 | 19903 | 37.311 | 19829 | 37.311 | 21406 |
| $HE_{16\,1\,1}$ | 37.750 | 42962 | 37.730 | 43116 | 37.730 | 43058 |
| $EH_{16\,1\,1}$ | 39.212 | 22782 | 39.248 | 22743 | 39.248 | 25713 |
| $HE_{17\,1\,1}$ | 39.578 | 45519 | 39.556 | 47444 | 39.556 | 45789 |

The same approach, in combination with linear approximation, was applied to parameters of the plastic using the model with the air-filled microfluidic channel. The parameters of the microfluidic channel have little effect on the resonator characteristics in the case of air filling. The permittivity and loss tangent values for different modes are shown in Fig. 5(b). The real part of permittivity changes from 2.26 to 2.24 and the loss tangent from 0.001 to 0.00064 in the frequency range from 30.16 GHz to 39.34 GHz.

A comparison of the simulated values using the above-mentioned plastic permittivity data and measured spectrum of the quartz resonator with air-filled microfluidic channel is shown in TABLE II.

### *A. Calibration of multifrequency setup based on a single WGM resonator*

For the calibration of the microwave setup, operating on six different modes, a procedure similar to the one previously suggested for investigations with a single Lorentzian-shaped mode [2],[15] was used for every mode. It is based on the extraction of the complex permittivity using a nomogram chart. It takes time to calculate the charts, but this calculation is performed only once prior to the experimental investigations and does not increase the measurement time.

TABLE II
EXPERIMENTAL AND SIMULATED SPECTRUM OF THE QUARTZ RESONATOR WITH AIR-FILLED MICROFLUIDIC CHANNEL

| Mode | Simulation | | Experiment | |
|---|---|---|---|---|
| | *f* (GHz) | *Q* | *f* (GHz) | *Q* |
| $HE_{12\,1\,1}$ | 30.165025 | 4444 | 30.165045 | 4445 |
| $HE_{13\,1\,1}$ | 31.999581 | 8490 | 31.999539 | 8528 |
| $HE_{14\,1\,1}$ | 33.835168 | 14064 | 33.835195 | 14462 |
| $HE_{15\,1\,1}$ | 35.672416 | 20064 | 35.672412 | 20457 |
| $HE_{16\,1\,1}$ | 37.510542 | 25452 | 37.511278 | 25473 |
| $HE_{17\,1\,1}$ | 39.351317 | 29946 | 39.351660 | 27537 |

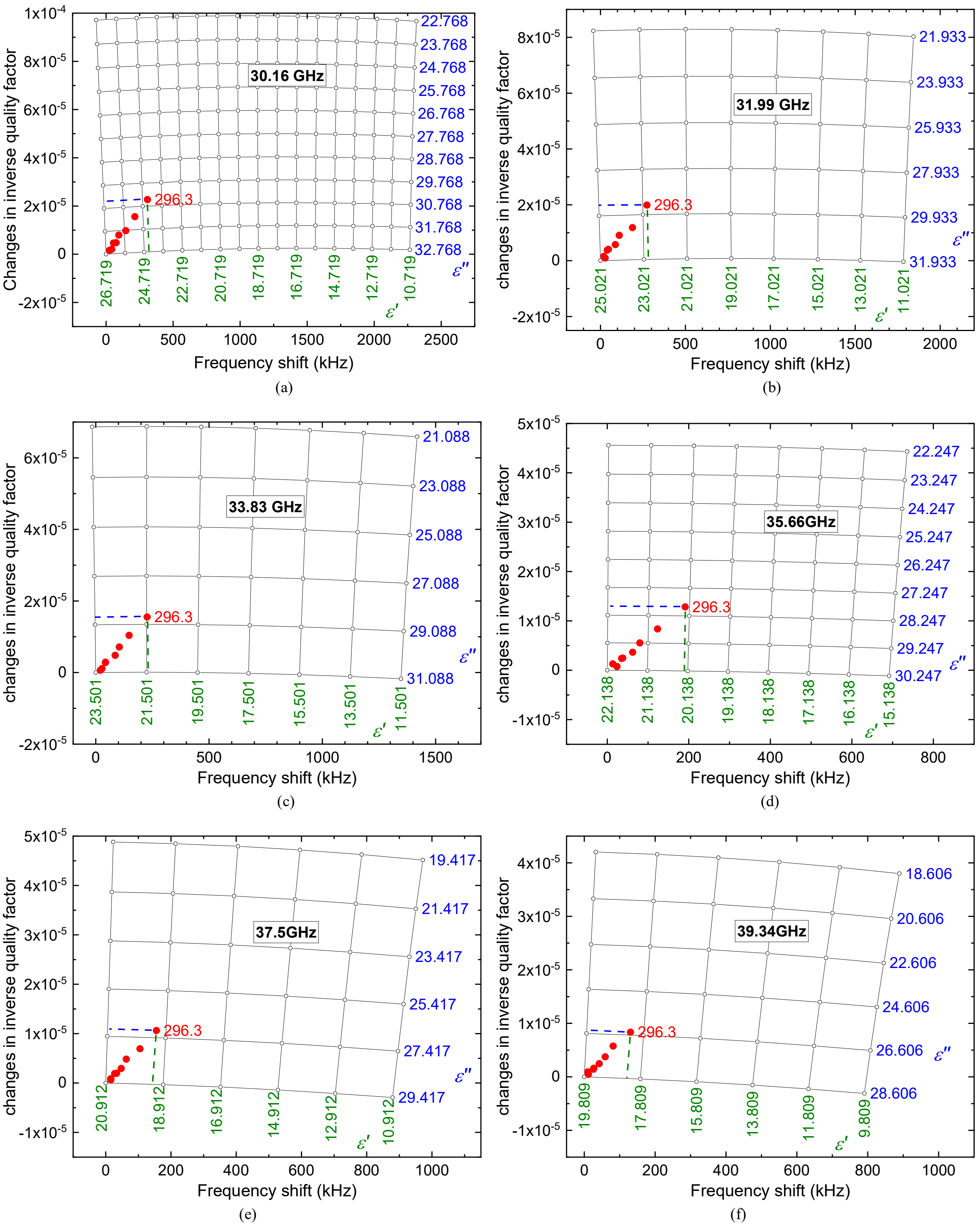

Fig. 6. *(Continued.)* Calibration nomogram charts. Calculated net (small open points) for the resonant frequency shift $\Delta f = f_{liq} - f_{water}$ and changes in the inverse quality factor $\Delta(1/Q) = 1/Q_{water} - 1/Q_{liq}$ for a microfluidic channel filled with substances with varying real and imaginary parts of permittivity obtained at six different frequencies: (a) 30.16 GHz ($HE_{12\,1\,1}$), (b) 31.99 GHz ($HE_{13\,1\,1}$), (c) 33.83 GHz ($HE_{14\,1\,1}$), (d) 35.66 GHz ($HE_{15\,1\,1}$), (e) 37.5 GHz ($HE_{16\,1\,1}$), and (f) 39.34 GHz ($HE_{17\,1\,1}$). The red solid points show the experimentally obtained results for eight different concentrations of L-lysine solutions in water (mmol/l): 17.3, 26.0, 39.0, 58.5, 87.8, 131.7, 197.5, and 296.3.

Each of the six nomogram charts is obtained by simulating the setup with some calibration liquid discrete changes of the real and imaginary parts of permittivity (open dots in Fig. 6).

The nomogram charts were obtained for all six measurement modes [Fig. 6(a) 30.16 GHz ($HE_{12\ 1\ 1}$), (b) 31.99 GHz ($HE_{13\ 1\ 1}$), (c) 33.83 GHz ($HE_{14\ 1\ 1}$), (d) 35.66 GHz ($HE_{15\ 1\ 1}$), (e) 37.5 GHz ($HE_{16\ 1\ 1}$), and (f) 39.34 GHz ($HE_{17\ 1\ 1}$)]. The difference from previous investigations is the need to maintain reliability for all measurement modes, i.e. simultaneous resonant frequencies. The experimentally obtained data for 17.3 mmol/l, 26 mmol/l, 39 mmol/l, 58.5 mmol/l, 87.8 mmol/l, 131.7 mmol/l, 197.5 mmol/l, and 296.3 mmol/l L-lysine concentration in water solutions are plotted in Fig. 6 as solid red dots.

The complex permittivity data for each concentration of L-lysine solution (red point) can be obtained from the chart using the graphical method. To extract the permittivity quickly, the Python code was written to obtain the value from linear equations from the four points that are closest to the experimental ones when combined.

## III. Liquid Permittivity Investigation of Six Modes Using Single WGM Resonator

The results of the complex permittivity investigation of lysine in water solution (17.3 mmol/l, 26.0 mmol/l, 39.0 mmol/l, 58.5 mmol/l, 87.8 mmol/l, 131.7 mmol/l, 197.5 mmol/l, and 296.3 mmol/l) as a function of frequency are shown in Fig. 7. The solutions were obtained from the maximum concentration solution (1 mol/l) from the L-lysine powder [27] dissolved in distilled water by dissolving two parts of the solutions in one part of distilled water for each of the solutions with a lower concentration.

The real part of permittivity dependence on frequency decreases monotonically with decreasing frequency for all of the investigated concentrations [Fig. 7(a)]. The imaginary part of the permittivity dependence has an almost linear dependence on frequency [Fig. 7(b)]. The permittivity value decreases with increasing lysine concentration in the water solution.

The reliability of the investigation of the $HE_{15\ 1\ 1}$ mode (35.6 GHz) was demonstrated by the measurements of known liquids like methanol, ethanol, propanol, and glucose aqueous solutions in [15]. The multifrequency investigation (the remaining five modes) should exhibit the same reliability, since the calibration was performed in a similar way. The reliability was additionally proven by means of comparison with the calculated data (shown in Fig.6; solid thick lines). The data were calculated by a sum of the symmetric Cole–Cole function corresponding to water and amino acid, and the conductivity term. The fitting parameters were determined in [21] by fitting the measured data of 0.24 mol/l L-lysine in water solution in a frequency range up to 20 GHz. The calculated curves, both for the real and imaginary permittivity parts, correspond to between the 197.5 mmol/l and 296.3 mmol/l curves obtained in our work and have the same frequency dependencies, despite the higher measurement frequencies. The measurement temperature (22°C) in [21] differs from the temperature in our study (25°C). The relaxation strength values were lowered taking account of the fact that the temperature is 3°C higher, which is also pointed out in [21].

The microfluidic chip, which was identical to the MFC used in [15], was optimized for investigations of aqueous solutions at a frequency of about 35.5 GHz in this case. Therefore, the maximum accuracy of the setup using the MFC of such geometrical dimensions was reached for a frequency of 35.5 GHz. The permittivity determination accuracy evaluated as $\delta\varepsilon'/\varepsilon'$, where $\delta\varepsilon'$ represents the most probable errors of the real part of permittivity, is equal to 1.2%, 0.9%, 0.7%, 0.5%, 0.4%, and 0.7% for modes where m=12, 13, 14, 15, 16, and 17 respectively. The value of $\delta\varepsilon''/\varepsilon''$, where $\delta\varepsilon''$ represents the most probable errors of the imaginary part of permittivity, is equal to 1.1%, 0.7%, 0.3%, 0.4%, 0.4%, and 0.6% for m=12, 13, 14, 15, 16, and 17 respectively. They were derived as a sum of the errors caused by the temperature variation of the measurement cell, the resonance parameters determined by fitting, and the absolute permittivity value determined from the nomogram chart. The first factor has the greatest influence on total error. Type A standard uncertainty calculated using [28] for the real part of permittivity is equal to 0.014, 0.013, 0.016, 0.018, 0.017, and 0.02; for the imaginary part of permittivity, it is equal to 0.01, 0.009, 0.006, 0.005, 0.005, and 0.008 for modes where m=12, 13, 14, 15, 16, and 17 respectively.

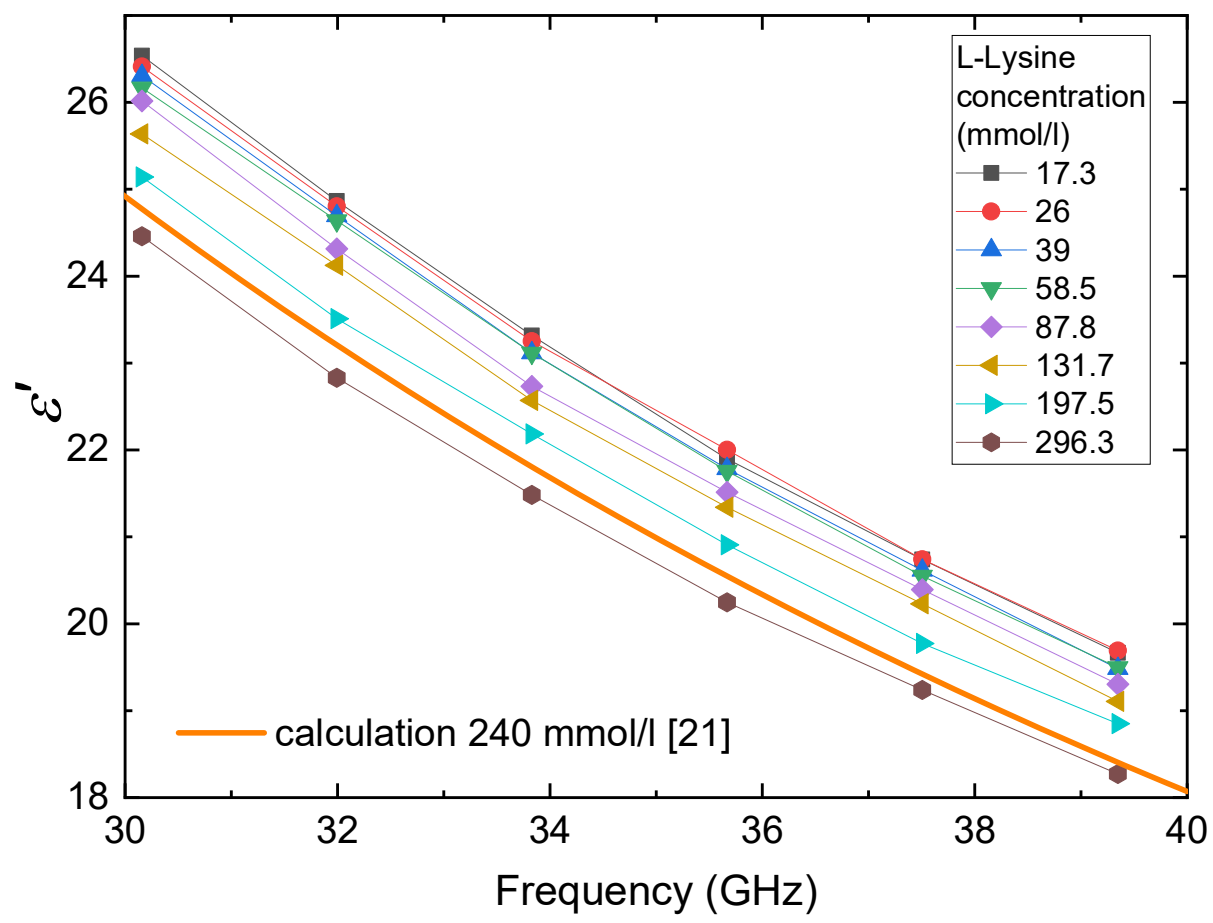

(a)

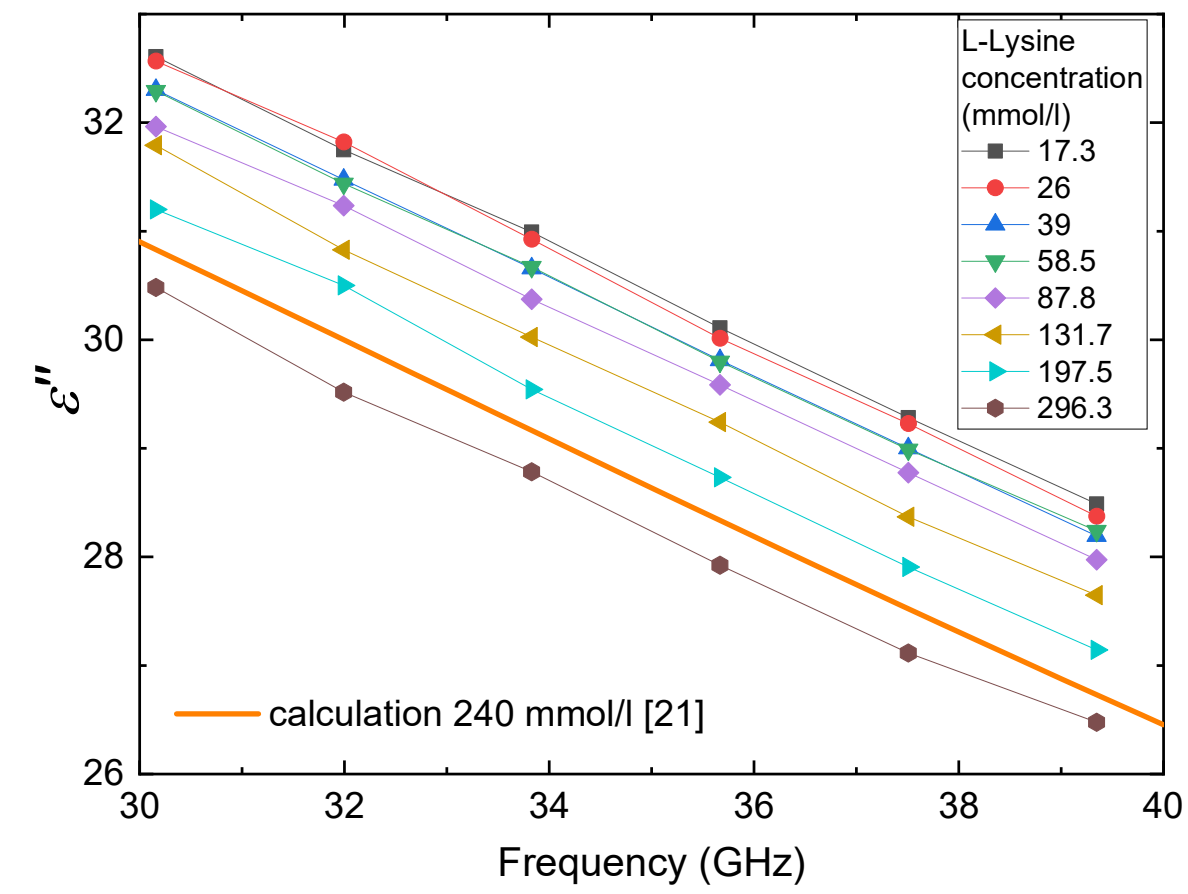

(b)

Fig. 7. Dependencies of the (a): real and (b): imaginary parts of lysine complex permittivity on frequency for different concentrations in water solutions (17.3 mmol/l, 26.0 mmol/l, 39.0 mmol/l, 58.5 mmol/l, 87.8 mmol/l, 131.7 mmol/l, 197.5 mmol/l, and 296.3 mmol/l). The calculation of 240 mmol/l lysine in water permittivity dependence by [21] is depicted by the thick solid line.

It should be emphasized that despite the fact that the MFC dimensions were optimized exclusively for an investigation at a frequency of about 35.5 GHz, the investigation results of lysine in water solutions for the other five frequencies also show high sensitivity.

## IV. Conclusion

In this work, the advanced technique based on a single WGM dielectric resonator with microfluidic chip to measure complex liquid permittivity at multiple frequencies was developed and demonstrated using the quasi-optical nature of the resonators. The sensor fabricated on the basis of crystal quartz material allows measurements to be performed at six discrete frequencies in the frequency range from 30 GHz to 40 GHz without changing the coupling. We demonstrate that the number of discrete measurement frequencies and their spacing can be tuned by changing the resonator dimensions. Sensor calibration was performed for each of the six modes. The resonance parameters measured for all symmetrical and non-symmetrical resonance responses were obtained using the coupled mode resonator fit. The complex permittivities of L-lysine solutions were obtained in a relatively wide frequency range of 10 GHz. The obtained results demonstrate a high level of accuracy and open up the possibilities for investigating biological liquids through the developed resonant method in a wide frequency range.

## References

[1] X. Jiang, A.J. Qavi, S.H. Huang, and L. Yang, “Whispering-Gallery Sensors,” *Matter*, vol.3, No 2, pp. 371–392, Aug. 2020.

[2] A.I. Gubin, A.A. Barannik, N.T. Cherpak, I.A. Protsenko, A. Offenhaeusser, and S. Vitusevich, “Whispering-gallery-mode resonator technique with microfluidic channel for permittivity measurement of liquids,” *IEEE Trans. Microw. Theory Techn.*, vol. 63, No 6, pp. 2003– 2009, Jun. 2015.

[3] Yan-Zhen Wei, and S. Sridhar, “Technique for measuring the frequency-dependent complex dielectric constants of liquids up to 20 GHz,” *Review of Scientific Instruments*, vol. 60, 3041, Jun. 1989.

[4] M. Pereira, and O. Shulika, “THz for CBRN and explosives detection and diagnosis,” in NATO Science for Peace and Security Series B: Physics and Biophysics. Amsterdam, The Netherlands: Springer, 2017, pp. 37–42.

[5] Katia Grenier, David Dubuc, Tong Chen, François Artis, Thomas Chretiennot, M. Poupot, and J-J. Fournié, “Recent Advances in Microwave-based Dielectric Spectroscopy at the Cellular Level for Cancer Investigations,” *IEEE Trans. Microw. Theory Techn.*, vol. 61, No 5, pp. 2023-2030, May 2013.

[6] “Keysight technologies basics of measuring the dielectric properties of materials,” Keysight Technol., Santa Rosa, CA, USA, Appl. Note 5989-2589EN, Jan. 2019.

[7] Emisens. Accessed: Dec. 2021. [Online]. Available: http://emisens.com.

[8] A.C. Stelson, M. Liu, C.A.E. Little, C.J. Long, N.D. Orloff, N. Stephanopoulos, and J.C. Booth, “Label-free detection of conformational changes in switchable DNA nanostructures with microwave microfluidics,” *Nat. Commun.*, vol. 10, 1174, Mar. 2019.

[9] Julie Lutti, Wolfgang Langbein, and Paola Borri, “A monolithic optical sensor based on whispering-gallery modes in polystyrene microspheres,” *Appl. Phys. Lett.,* vol. 93, 151103, Oct. 2008.

[10] Martin D. Baaske, Matthew R. Foreman, and Frank Vollmer, “Single-molecule nucleic acid interactions monitored on a label-free microcavity biosensor platform,” *Nature Nanotech.*, vol. 9, pp. 933–939, Aug. 2014.

[11] A.A. Barannik, N.T. Cherpak, Yu.V. Prokopenko, Yu.F. Filipov, E.N. Shaforost, and I.A. Shipilova, "Two-layered disc quasi-optical dielectric resonators: electrodynamics and application perspectives for complex permittivity measurements of lossy liquids," *Meas. Sci. Technol.* vol. 18, pp. 2231-2238, Jul. 2007.

[12] O.N. Shaforost, N. Klein, S.A. Vitusevich, A.A. Barannik, and N.T. Cherpak, “High sensitive microwave characterisation of organic molecule solutions of nanolitre volume,” *Appl. Phys. Lett.*, vol. 94, No 11, p. 112901, Mar. 2009.

[13] A. Gubin, A. Lavrinovich, I. Protsenko, A. Barannik, and S.Vitusevich, “WGM dielectric resonator with capillary for microwave characterization of liquids,” *Telecommunications and Radio Engineering*, vol. 78, No 18, pp.1651-1657, Dec. 2019.

[14] N. Cherpak, A. Lavrinovich, and E. Shaforost, “Quasi-optical dielectric resonators with small cuvette and capillary filled with ethanol-water mixtures,” *International Journal of Infrared and Millimeter Waves,* vol. 27, pp. 115-133, Nov. 2006.

[15] Alexey I. Gubin, Irina A. Protsenko, Alexander A. Barannik, Svetlana Vitusevich, Alexandr A. Lavrinovich, and Nickolay T. Cherpak, “Quartz Whispering-Gallery-Mode Resonator With Microfluidic Chip as Sensor for Permittivity Measurement of Liquids,” *IEEE Sensors Journal*, vol. 19, No. 18, Sep. 2019.

[16] Xiao Ma, Xiaotian Du, Lei Li, Caroline Ladegard, Xuanhong Cheng, and James C. M. Hwang, “Broadband Electrical Sensing of a Live Biological Cell with In Situ Single-Connection Calibration,” *Sensors*, vol. 20, 3844, Jul. 2020.

[17] Katia Grenier, Amar Tamra, Amel Zedek, Guillaume Poiroux, François Artis, Tong Chen, Wenli Chen, M Poupot, J-J Fournié, and David Dubuc, “Low Volume and Label-Free Molecules Characterization and Cell Monitoring with Microwave Dielectric Spectroscopy,” IEEE International Microwave Bio Conference (IMBioC 2018), Philadelphie, PA, United States, Jun. 2018.

[18] Xiang Zhao, Zhuo Cheng, Ming Zhu, Tianye Huang, Shuwen Zeng, Jianxing Pan, Chaolong Song, Yuhan Wang, and Perry Ping Shum, "Study on the dual-Fano resonance generation and its potential for self-calibrated sensing," *Opt. Express,* vol. 28, No. 16, pp. 23703-23716, Aug. 2020.

[19] COMSOL Multiphysics. Accessed: Jul. 2021. [Online]. Available: http://www.comsol.com.

[20] M. Zhadobov, R. Augustine, R. Sauleau, S. Alekseev, A. Di Paola, C. Le Quément, Y.S. Mahamoud, and Y. Le Dréan, “Complex permittivity of representative biological solutions in the 2-67 GHz range,” *Bioelectromagnetics,* vol. 33, No. 4, pp. 346-55, May 2012.

[21] Iñigo Rodríguez-Arteche, Silvina Cerveny, Ángel Alegría, and Juan Colmenero, “Dielectric spectroscopy in the GHz region on fully hydrated zwitterionic amino acids,” *Phys. Chem. Chem. Phys.*, vol. 14, pp. 11352–11362, Jun. 2012.

[22] A. I. Gubin, I. A. Protsenko, A. A. Barannik, H. Hlukhova, N. T. Cherpak, and S. Vitusevich, "Liquids Microwave Characterization Technique Based on Quartz WGM Resonator with Microfluidic Chip," Proc. of 48rd Eur. Microwave Conf., Sep. 23-28, Madrid, Spain, 2018, pp. 206-209.

[23] Zeonex. Accessed: Dec. 2021. [Online]. Available: https://www.zeonex.com.

[24] A. Ya. Kirichenko, Yu. V. Prokopenko, Yu. F. Filippov and N.T. Cherpak, Kvazioptičheskie tverdotelnye rezonatory (Quasioptical Solid-state Resonators). Kyiv, Ukraine: Naukova Dumka, 2008, p. 296.

[25] V.N. Skresanov, V.V. Glamazdin, and N.T. Cherpak, “The novel approach to coupled mode parameters recovery from microwave resonator amplitude-frequency response,” in Proc. of the 41rd European Microwave Conference EuMC’11, Oct. 10-13, Manchester, UK, 2011, pp. 826-829.

[26] J. Krupka, K. Derzakowski, M. Tobar, J. Hartnett and R. G. Geyerk, “Complex permittivity of some ultralow loss dielectric crystals at cryogenic temperatures,” Meas. Sc. Tech., vol. 10, No 5, pp. 387–392, May 1999.

[27] Sigma Aldrich. Accessed: Dec. 2021. [Online]. Available: https://www.sigmaaldrich.com.

[28] “Evaluation of measurement data — Guide to the expression of uncertainty in measurement”, JCGM 100:2008, Sep. 2008.

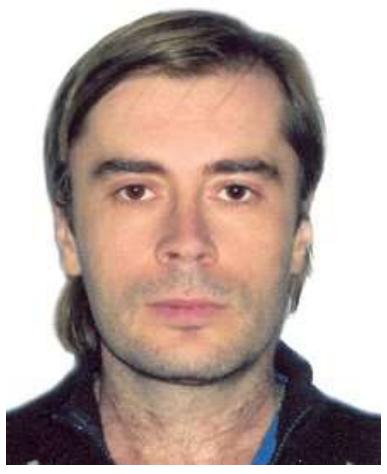

**Alexey I. Gubin** was born in Kharkiv, Ukraine, in 1978. He received a diploma in radiophysics and electronics from V.N. Karazin Kharkiv National University, Kharkiv, Ukraine, in 2000 and a Ph.D. (physical and mathematical sciences) in radiophysics from O. Ya. Usikov Institute for Radiophysics and Electronics, National Academy of Science of Ukraine, Kharkiv, Ukraine, in 2007.

Since 2000, he has been working at the O. Ya. Usikov Institute for Radiophysics and Electronics, National Academy of Science of Ukraine, Kharkiv, Ukraine. Currently, he is a Senior Researcher. In 2017, he became Associate Professor. His scientific activities encompass investigations of the microwave to submillimeter-wave characteristics of condensed matter using techniques based on the non-resonant grazing incidence reflectivity technique and the whispering-gallery-mode dielectric resonator based technique.

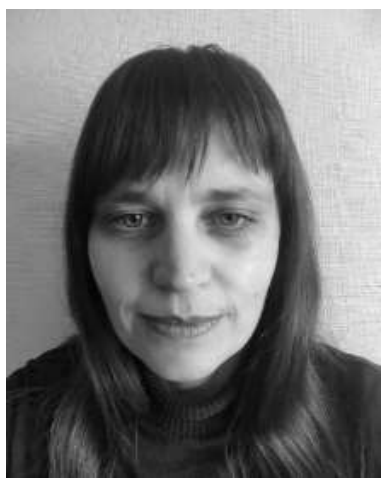

**Irina A. Protsenko** received a diploma in radio engineering from Kharkiv National University of Radio Electronics in 2005 and a Ph.D. (physical and mathematical sciences) in radiophysics from O. Ya. Usikov Institute for Radiophysics and Electronics, National Academy of Science of Ukraine, Kharkiv, Ukraine, in 2018.

She has been working at the Institute of Radiophysics and Electronics, National Academy of Science of Ukraine, Kharkiv, Ukraine since 2005. Her current research interests include measuring the electrophysical parameters of materials.

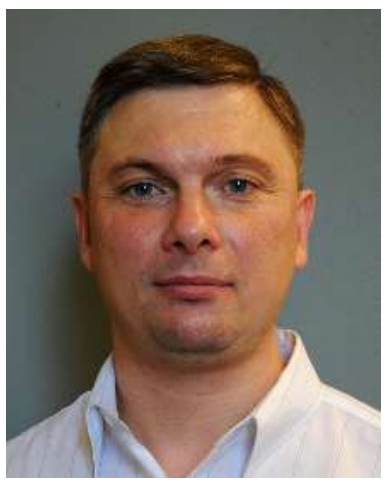

**Alexander A. Barannik** was born in Kharkiv, Ukraine, on April 23, 1975. He received a diploma in cryogenic techniques and technology from the National Technical University "Kharkiv Polytechnic Institute", Kharkiv, Ukraine, in 2000 and a Ph.D. (physical and mathematical sciences) in radiophysics from Kharkiv National University of Radio Electronics, Kharkiv, Ukraine in 2004.

Since 2000, he has been working at the O. Ya. Usikov Institute for Radiophysics and Electronics, National Academy of Science of Ukraine, Kharkiv, Ukraine, as a Junior Researcher (2000), Scientific Researcher (2005), and Senior Researcher (2007). He has co-authored over 100 scientific publications and he has five patents. His scientific activities focus on studying the microwave characteristics of condensed matter, including HTS, dielectrics, and liquids using whispering-gallery-mode dielectric resonators. He also studies the electrodynamic properties of various types of whispering-gallery-mode resonators.

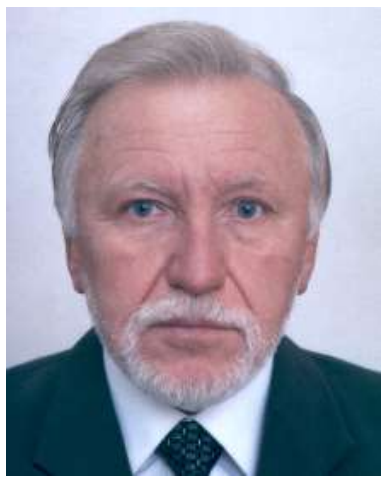

**Nickolay T. Cherpak** (Senior Member, IEEE) received D.Sc. degrees in 1987. He is a Chief Researcher at the O. Ya. Usikov Institute for Radiophysics and Electronics, National Academy of Sciences of Ukraine (NASU), Kharkiv, Ukraine. He was previously engaged in the development of low-noise solid-state microwave and millimeter (mm)-wave maser amplifiers. At present, he is the head of a research group that is developing the mm-wave measurement technique for the characterization of condensed matter such as unconventional superconductors and lossy dielectric liquids. The group also performs the relevant physical studies.

In 1999, Prof. Cherpak received the I. Puljuy Award of Presidium of NASU for the best experimental work on RF and microwave studies of HTS. He also received the Academician Sinelnikov Award from the Kharkiv Regional Administration (2013) and the L. Shubnikov Award from the Presidium of NASU for the best experimental work on microwave impedance studies of new, unconventional superconductors (2016). Prof. Cherpak is the author and co-author of about 370 publications, including three books, five reviews, and over 20 patents (Former SU, Ukraine and USA). He has also supervised seven Ph.D. dissertations and was a Professor at the National Technical University (NTU) – Kharkov Polytechnic Institute for more than 20 years. Currently, he is a member of the Specialized Council for the award of the scientific degree of Doctor of Science at the National University of Radio Electronics in Kharkiv. He is an editorial staff member of two scientific journals. Prof. Cherpak is a member of URSI Commission E in Ukraine and was a Chairman of the IEEE Joint East Ukraine Chapter (2014-2017). He has been chairman, program committee member, and session chair of a number of international conferences and workshops.

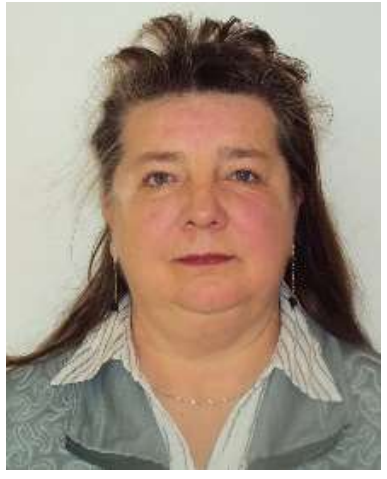

**Svetlana Vitusevich** received an M.Sc. degree in radiophysics and electronics from Kiev State University, Kiev, Ukraine, in 1981, a Ph.D. degree in physics and mathematics from the Institute of Semiconductor Physics (ISP), Kiev, Ukraine, in 1991, and a Dr. habil. degree from the Supreme Attestation Commission of the Ukraine, Kiev, Ukraine, and Technical University of Dortmund, Dortmund, Germany, in 2006.

From 1981 to 1997, she worked at the ISP as a Researcher (1981), Scientific Researcher (1992), and Senior Scientific Researcher (1994). From 1997 to 1999, she worked at the Institute of Thin Films and Interfaces (ISI), Forschungszentrum Jülich (FZJ), Jülich, Germany, as an Alexander von Humboldt Research Fellow. Since 1999, she has been a Senior Scientific Researcher with the Institute of Complex Systems – Bioelectronics (ICS-8, formerly Peter Grünberg Institute – PGI-8, IBN-2, ISG-2, and ISI), FZJ. In 2006, she became the leader of a working group whose main topic is dedicated to 1/f noise in quantum heterostructures and its up-conversion into phase noise in high-frequency oscillators, including high-quality resonators. She is an active teacher with the Technical University of Dortmund, Dortmund, Germany. She has authored or co-authored over 170 papers in peer-reviewed scientific journals. She holds ten patents, among them a dual-mode resonator for the simultaneous measurement of dielectric relaxation and the electrical conductivity of liquids in closed bottles. Her research interests are focusing on transport and noise properties of different kinds of materials for the development of label-free biosensors and novel device structures for future information technologies.